\documentclass[11pt]{article}

\usepackage{amssymb,latexsym}
\usepackage{amsmath}
\usepackage{amsfonts}
\usepackage{amsopn}
\usepackage{amsthm}
\usepackage{epsfig}
\usepackage{lscape}
\usepackage{multirow}
\usepackage{url}
\usepackage[pdftex,bookmarks=true]{hyperref}
\usepackage{pgf,tikz}
\usetikzlibrary{arrows}
\evensidemargin\oddsidemargin \topmargin-15mm

\newcounter{NN}
\newtheorem{remark}[NN]{Remark}

\newtheorem{definition}[NN]{Definition}

\def\Z{\mathbb{Z}}

\def\Z{\mathbb{Z}}

\begin{document}
\title{A Lax pair  of a lattice equation whose entropy vanishes}
\author{Dinh T. Tran\\
\\Department of Mathematics and Statistics,
\\
La Trobe University, Bundoora VIC 3086, Australia\\
Email:  Dinh.tran@latrobe.edu.au or dinhtran82@yahoo.com}
\maketitle
\maketitle
\begin{abstract}
In this paper, we present multi parametric  quadgraph equations which are consistent around the cube. These equations are obtained by applying a `double twist' to known integrable equations. Furthermore, we perform a limit to one of these equations to derive the non-symmetric  equation which is Equation 19 in \cite{HV}. As a result, we obtain a novel  Lax pair of this equation.  \end{abstract}
 \section{Introduction}
 There are various definitions of integrability for both continuous and discrete integrable systems. One of the common definitions is  the notion of having a Lax representation. 
 Integrable non-linear equations arise as the compatibility condition of two linear systems called a Lax pair.  
 The origins lie in the inverse scattering method introduced by Gardner, Greene, Kruskal and Miura \cite{Gardner} to solve the initial value problem for Korteweg-de Vries (KdV) equation.
In this paper, we consider only lattice integrable equations on a square 
\begin{equation}
\label{E:PDeltaE}
Q(u_{l,m}, u_{l+1,m},u_{l,m+1},u_{l+1,m+1})=0,
\end{equation}
whose  Lax pair is often given by a  pair of  matrices. In this equation,  the field variable, $u$, depends on lattice variables $l$ and  $m$ where $(l,m)\in \Z^2$.

The definition of integrability associated with having a Lax pair is quite simple. Nevertheless, the question arising is how to construct a Lax pair of a given equation. There is a systematic way of constructing 
a Lax pair of equations on quad graphs that satisfy consistency around the cube (CAC) and other properties \cite{ABS,  Nijhoff, NW}.  The CAC property gives us a Lax pair for the equations directly. 
There is a list of Lax pairs for many  CAC equations and systems,  including equations in the Adler-Bobenko-Suris (ABS) list \cite{ABS, ABS2} given in \cite{BHQK}.
However, in general if a lattice equation  is not CAC, it would be difficult to find its Lax pair if it exists. 
 
Integrable systems are often associated with `low complexity'.  One of the means to measure the complexity of a discrete system is called algebraic entropy \cite{howtodetect, TGR, LatticeEntropy}. 
It is believed that discrete  integrable mappings and lattice equations have vanishing entropy. In other words, given a map  we can use initial values to iterate the map.  For an integrable map, one would get a polynomial growth of degrees of the map. It also holds for integrable  lattice equations. Suppose that we can solve uniquely at any vertex and obtain a rule  to propagate new vertices. 
Vanishing entropy  can be seen  through factorisation and  large cancellations.   As a consequence, it gives a slow growth of degrees of iterations.  Vanishing entropy for lattice equations has been used as a simple integrability test for 
a decade. However, it has only been proven  recently for certain lattice equations subject to a conjecture~\cite{RobertsTran}.

On the other hand, `low complexity`  has been used in   \cite{HV} to search for  integrable lattice equations on a square. The authors searched for multi-affine  equations on a quad-graph with a certain factorisation property. Two of the highlights in their paper  are   Equation 19  and Equation 25. These equation are given as follows
\begin{equation}
\label{E:19}
{\rm E19}:\ (x-v)(u-y)+r_4(x-y)+r_2(u-v)+s=0,
\end{equation}
and
\begin{equation}
\label{E:25}
{\rm E25}:\
xy+uv+p_3(xu+vy)-(p_3+1)(xv+uy)+r_4(y-x)+r_2(u-v)
-\left(s(p_3+1)+r_4\right)(p_3s+r_4)+ s r_2 
=0,
\end{equation} 
where we have  used the  short notation given in \cite{ABS}, i.e.  we denote $u_{l,m}=x, u_{l+1,m}=u, u_{l,m+1}=v$ and $u_{l+1,m+1}=y$.
These two  equations have  vanishing entropy;  therefore, they are likely to be integrable.   
 Equation E25  given by~\eqref{E:25} is  related to $Q_1$ via a non-autonomous transformation except for $p_3=0$ or $p_3=-1$. 
However, equation  E19 (see equation~\eqref{E:19})  seems to be new as it is not symmetric. 
This makes it hard to find its Lax pair (if it exists) as we can not put  this equation on a cube. 
So far, not much has been known about this equation. Furthermore, 
at the first glance, we can see that equation E19 is a special case of equation E25. In fact, it falls into an exceptional case of the latter one where $p_3=-1$. Thus,  it seems that it would be  hard to connect $Q_1$  and equation E19  directly but
one still hopes to establish connections  through  a limit procedure.

 Recently, there have been  interesting results obtained  by Ormerod et al in \cite{Twisted}. The authors introduced the notion of a  twist matrix which is associated with a twisted  reduction $T$ where $T$ is taken as a point symmetry of quad-graph equations. If we could find a twist matrix corresponding to $T$, we will obtain Lax pairs of twisted reductions which are generalisation of periodic reductions. In this paper, we use this idea at  the level of lattice equations .   In fact, the twist  acts only on two vertices on one edge of an equation on a square. It gives different transformations on four points of a square. In general, those transformations do not preserve nice properties of integrable equations such as Lax pairs and vanishing entropy.  
 
 We note that a non-autonomous transformation sometimes can be written as a twist. Therefore, it suggests that we might be able to connect $Q_1$ with equation E25 via some twists.
In this paper, we establish a connection between these two equations. This will help to find autonomous  Lax pairs for equations E25 and E19. 

This paper is organised as follows. In section 2, we will set up some definitions  that will be needed for the rest of the paper. In the  next section, we will present some CAC equations (including  equation E25) with multi-parameters.  These equations are obtained through a `double twist'. We note that the CAC property provides us with a Lax pair. However, the double twist also gives us a Lax pair for the corresponding equation.   In particular,  we present a Lax pair of equation E25 with a different parametrisation. It has been done through two twists of $Q_1$. In section 4, we present in detail how to derive a Lax pair for equation E19 from the  Lax pair of equation E25.

\section{The setting}
In this section, we introduce  some notions that will be used in this paper. 
We consider equation~\eqref{E:PDeltaE} on a quad graph.
The  equation is called integrable if it arises as  the compatibility condition of the following system of linear equations
\begin{equation}
\label{E:PDeltaECompatibility}
\begin{array}{ll}
\phi_{l+1,m}&=L_{l,m}\phi_{l,m}, \\
\phi_{l,m+1}&=M_{l,m}\phi_{l,m},
\end{array}
\end{equation}
where $\phi$ is a vector,  $L$ and  $M$ are matrices that depend on spectral parameter, $\lambda$. Generally $L$
 depends on $u_{l,m}, u_{l+1,m}$ and $M$ depends on $u_{l,m}, u_{l,m+1}$.
Thus, equation~\eqref{E:PDeltaE}  is  integrable if it satisfies  the compatibility condition
\begin{equation}
\label{E:LaxPDeltaE}
L_{l,m+1}M_{l,m}-M_{l+1,m}L_{l,m}=0.
\end{equation}
\begin{definition}
\label{D:Lax}
A pair of matrices, $L$ and $M$, satisfying~\eqref{E:LaxPDeltaE}  if  equation~\eqref{E:PDeltaE} satisfied is called a Lax pair of equation~\eqref{E:PDeltaE}.
\end{definition}
For example, equation $Q_1$ which is given  by cf.\cite{ABS}
\begin{equation}
\label{E:Q1}
p(x-v)(u-y)-q(x-u)(v-y)+\delta^2pq(p-q)=0
\end{equation}
has the following Lax pair cf. \cite{BHQK}
\begin{equation}
\label{E:Q1_Lax}
L(x,u,p):=\begin{pmatrix}
{\frac {-ku+kx+pu}{\delta\,p-u+x}}&{\frac {
{\delta}^{2}p{k}^{2}-{\delta}^{2}{p}^{2}k-pxu}{\delta\,p-u+x}}
\\ 
{\frac {p}{\delta\,p-u+x}}&{\frac {-ku+kx-px}{\delta\,p-u+x}}
\end{pmatrix},\
M(x,v,q):=\begin{pmatrix}
{\frac {-kv+kx+qv}{\delta\,q-v+x}}&{\frac {
{\delta}^{2}q{k}^{2}-{\delta}^{2}{q}^{2}k-qxv}{\delta\,q-v+x}}
\\ 
{\frac {q}{\delta\,q-v+x}}&{\frac {-kv+kx-qx}{\delta\,q-v+x}}
\end{pmatrix}.
\end{equation}
We perform a transformation, $\phi_{l,m}\rightarrow G_{l,m} \phi_{l,m}$, where $G$ is any non-singular matrix. This gives $\widetilde{L}_{l,m}=G_{l+1,m}L_{l,m}G_{l,m}^{-1}$ and
 $\widetilde{M}_{l,m}=G_{l,m+1}M_{l,m}G_{l,m}^{-1}$, and the equation~\eqref{E:PDeltaE} is invariant under this transformation. We call this transformation a gauge transformation and $G$ is a 
gauge matrix. In fact,  it just  gives us a different form of a Lax representation for equation~\eqref{E:PDeltaE}.

%\begin{definition}
Recall that a  transformation $T$ is called a point symmetry of ~\eqref{E:PDeltaE} if
\begin{equation}
\label{E:symmetry}
Q\left(T(u_{l,m}),T(u_{l+1,m}),T(u_{l,m+1}),T(u_{l+1,m+1})\right)=0.
\end{equation}
%\end{definition}
In \cite{Twisted}, the authors introduced the quasi-periodicity $u_{l+s_1,m+s_2}=T(u_{l,m})$ as the $(s_1,s_2)$ twisted reduction.
A direct method to obtain a Lax pair of these reductions was  presented.
Given an integrable equation $Q$ with a Lax pair $(L,M)$, we consider the following equation
\begin{equation}
\label{E:Twisted_Eq}
Q\left(T(u_{l,m}),T(u_{l+1,m}),u_{l+1,m},u_{l+1,m+1}; p,q\right)=0,
\end{equation}
where  $T$  is a  point transformation such that the above equation 
has a vanishing entropy. It is noted that $T$ does not need to be a point symmetry of $Q$.
We have
\[
M\left(T(u_{l+1,m})u_{l+1,m+1},q\right)L\left(T(u_{l,m}),T(u_{l+1,m},p)\right)=L\left(u_{l,m+1},u_{l+1,m+1},p\right) M\left(T(u_{l,m}),u_{l,m+1},q\right).
\]
We look for a twist  matrix $S$ that satisfies $$L\left(T(u_{l,m}),T(u_{l+1,m},p)\right) S(u_{l,m},p)=S(u_{l+1,m},p)L(u_{l,m},u_{l+1,m},p).$$
Therefore, we have
\begin{multline*}
M\left(T(u_{l+1,m})u_{l+1,m+1},q\right)S(u_{l+1,m},p)L(u_{l,m},u_{l+1,m},p)S^{-1}(u_{l,m},q)=\\
L\left(u_{l,m+1},u_{l+1,m+1},p\right) M\left(T(u_{l,m}),u_{l,m+1},q\right).
\end{multline*}
This implies that $\left(L(u_{l,m},u_{l+1,m},p),M\left(T(u_{l,m}),u_{l,m+1},q\right)S(u_{l,m},p)\right)$ is a Lax pair for \eqref{E:Twisted_Eq}.
We note that  for many cases,  twist matrices might not exist.

%**************************************************************************************************************************************************************************************************
\section{CAC equations with multi-parameters}

In this section,  we present some equations which are CAC. These equations are obtained by performing a double twist on $Q_1$ and $H_1$. Using a twist matrix we obtain Lax pairs for these equations.  We use $Q_1$ as an example. We present a Lax pair for equation  E25  (equation \eqref{E:25})  through a connection with $Q_1$. 
As mentioned above,  when $p_3\neq 0,-1$, one can bring this equation to $Q_1$ via a transformation that depends on lattice variables $l$ and $m$.
Therefore, we can obtain  a non-autonomous Lax pair of this equation from  a Lax pair of $Q_1$. However, our aim is to find a Lax pair that does not depend on  $l$ and $m$
explicitly.

We first rewrite  equation E25 with a different parametrisation  as follows
\begin{equation}
\label{E:Dtwist}
p(x-v+c_1)(u-y+c_1)-q(x-u+c_2)(v-y+c_2)+\delta^2 pq(p-q)=0.
\end{equation}
This equation is a `double twist' of $Q_1$ in the sense that 
it is given by
\begin{equation}
\label{E:DT}
Q\left(T_2(T_1(x)),T_1(u),T_2(v),y,p,q\right)=0,
\end{equation}
where  $T_1(x)=x+c_1$ and $T_2(x)=x+c_2$. 

For a single twist $T_1$,  we first  look for a constant twist matrix $S_1$. By solving the equation
$L(T_1(x),T_1(u),p)S=S L(x,u,p)$, where $L$ is given by \eqref{E:Q1_Lax}, one gets
\begin{equation}
S_1=
\begin{pmatrix}
1& c_1\\
0&1
\end{pmatrix}.
\end{equation}
Therefore, we obtain a new $\mathcal{M}$ matrix of the equation $Q\left(T_1(x),T_1(u),v,y,p,q\right)=0$, where
\begin{equation}
\label{E:newM}
\mathcal{M}=
\left( \begin {array}{cc} {\frac {qv-kv+kx+kc_{{1}}}{\delta\,q-v+x+c_
{{1}}}}&{\frac {-{\delta}^{2}{q}^{2}k+{\delta}^{2}q{k}^{2}-qvx-kvc_{{1
}}+kxc_{{1}}+k{c_{{1}}}^{2}}{\delta\,q-v+x+c_{{1}}}}
\\ \noalign{\medskip}{\frac {q}{\delta\,q-v+x+c_{{1}}}}&{\frac {-qx-kv
+kx+kc_{{1}}}{\delta\,q-v+x+c_{{1}}}}\end {array} \right).
\end{equation}
One can then apply the second twist $T_2$ to find a new Lax matrix  $\mathcal{L}$. However, we note that the twists $T_1$ and $T_2$ are commutativeß,  i.e. $T_1(T_2(x))=T_2(T_1(x))$. 
Therefore, we can also apply the twist $T_2$ first. This gives us a new matrix  $\mathcal{L}$ for $Q\left(T_2(x),T_2(u),v,y,p,q\right)=0$, where 
\begin{equation}
\label{E:newL}
\mathcal{L}=
\left( \begin {array}{cc} {\frac {pu-ku+kx+kc_{{2}}}{\delta\,p-u+x+c_
{{2}}}}&{\frac {-{\delta}^{2}{p}^{2}k+{\delta}^{2}p{k}^{2}-pxu-kuc_{{2
}}+kxc_{{2}}+k{c_{{2}}}^{2}}{\delta\,p-u+x+c_{{2}}}}
\\ \noalign{\medskip}{\frac {p}{\delta\,p-u+x+c_{{2}}}}&{\frac {-px-ku
+kx+kc_{{2}}}{\delta\,p-u+x+c_{{2}}}}\end {array} \right).
\end{equation}
It is easy to check that $(\mathcal{L},\mathcal{M})$ is a Lax pair  for equation~\eqref{E:Dtwist}.
We note that  equation \eqref{E:Dtwist} is  consistent around the cube  with multi-parameters $\left((p,c_2),(q,c_1)\right)$. Thus, we can obtain a Lax pair directly from CAC. It actually gives us a Lax pair with two spectral parameters.
By taking one of these two spectral parameters to be $0$, we obtain the Lax pair $(\mathcal{L},\mathcal{M})$ given by~\eqref{E:newL} and~\eqref{E:newM}.

\begin{remark}
We give here two more  CAC equations.
\begin{itemize}
\item Similarly, when $\delta=0$, one can use two twists $T_3(x)=\alpha_1 x$ and $T_4(x)=\beta_1x$ on $Q_1$.  A twist matrix is given by
\begin{equation}
\label{E:Cross_ratio_twist}
S_{\alpha}=
\begin{pmatrix}
\alpha&0\\
0&1
\end{pmatrix}.
\end{equation}
This gives us a Lax pair of a mullti-parametric cross-ratio equation cf.~\cite{KT}
\begin{equation}
\label{E:Multi-cross_ratio}
p\beta_1(\alpha_1x-v)(\alpha_1 u-y)=q\alpha_1(\beta_1 x-u)(\beta_1v-y).
\end{equation}
This equation satisfies the 3D consistency and tetrahedron properties \cite{ABS} and has two  parameters in each direction. Moreover, it is actually a special case of Equation 16 in \cite{HV}.
\item Applying the twist $T_1$ and $T_2$ to $H_1$, we obtain the twist matrix
\begin{equation}
\label{E:H1_twist}
S_{c}=
\begin{pmatrix}
1&c\\
0&1
\end{pmatrix},
\end{equation}
and the following equation
\begin{equation}
\label{E:general_H1}
(x-y)(u-v)+(c_1+c_2)(u-v)+(c_1-c_2)(x-y)+c_1^2-c_2^2-p+q=0.
\end{equation}
This equation is a slightly more general form of Equation $B_1$ given in \cite{Field}.  We have checked that this equation is consistent around the cube with lattice parameters $\left((p,c_2), (q,c_1)\right)$.
\end{itemize}
\end{remark}

\begin{remark}
We note that all the equations which  we obtain in this sections can be transformed to $Q_1$ and $H_1$ (respectively) via  non-autonomous transformations.
\end{remark}
%---------------------------------------------------------------------------------------------------------------------------------------------------------------------------------------------------------------------------
\section{Lax pair of Equation 19}
In this section, we derive a Lax pair of Equation 19  given by~\eqref{E:19} from the previous  Lax pair of equation~\eqref{E:Dtwist}.

As discussed in the introduction, equation E19 is a special case of equation E25 with $p_3=-1$. In the new form \eqref{E:Dtwist} of equation E25 if we take $q=0$,
 we can get rid of the term $(x-u)(v-y)$ but the equation~\eqref{E:Dtwist} will  become trivial.  Hence, one needs to
avoid this problem by considering equation \eqref{E:19} as a limit of equation \eqref{E:Dtwist}.

The process of deriving  a Lax pair of this equation is given as follows.
\begin{itemize}
\item By dividing  both sides of the equation~\eqref{E:Dtwist} by $pq(p-q)$, we obtain the following equation
\begin{equation}
\label{E:Dtwist1}
\frac{(x-v+c_1)(u-y+c_1)}{q(p-q)}-\frac{(x-u+c_2)(v-y+c_2)}{p(p-q)}+\delta^2=0.
\end{equation}
\item In order to get  the term $(x-v)(u-y)$ and cancel the term $(x-u)(v-y)$, we choose $p,q$ such that $q(p-q)\rightarrow1$ and $p(p-q)\rightarrow \infty$. 
Thus, one can choose $q=\epsilon$ and $p=\epsilon+1/\epsilon$. To obtain the linear and constant  terms in the equation~\eqref{E:19}, we need to choose $c_2$ such that
$\frac{c_2}{p(p-q)}$ and $\delta^2-\frac{c_2^2}{p(p-q)}$ approach  to constants as $\epsilon \rightarrow 0$.
Therefore, we choose $c_2=\beta/\epsilon^2$ and $\delta=\beta \sqrt{s\epsilon^2+1}/\epsilon$. 
\item Substituting  these values into $\mathcal{L}$ and $\mathcal{M}$ given by \eqref{E:newL} and \eqref{E:newM}, one would hope  to obtain a Lax pair for equation E19 
 by taking the limit of these two matrices. However, there is a factor $\epsilon$ in a denominator of $\mathcal{L}_{12}$.  To avoid this, we choose a spectral parameter to be $\epsilon k$.  
We then divide  the matrices $\mathcal{L}, \mathcal{M} $ by $\epsilon$ to cancel the factor $\epsilon$ and take the limit of these results as $\epsilon\rightarrow 0$.  We obtain the following matrices
\begin{equation}
\label{E:Lax19L}
L_1=
 \left( \begin {array}{cc} \,{\frac {\beta\,k+u}{2\beta}}&\,{
\frac {{\beta}^{2}{k}^{2}-{\beta}^{2}ks-2\,{\beta}^{2}k-\beta\,ku+
\beta\,kx-ux}{2\beta}}\\ \noalign{\medskip}\frac{1}{2\beta}&\,{
\frac {\beta\,k-x}{2\beta}}\end {array} \right)
\end{equation}
and
\begin{equation}
\label{E:Lax19M}
M_1=
\left( \begin {array}{cc} -{\frac {kv-kx-kc_{{1}}-v}{\beta+c_{{1}}-v+
x}}&{\frac {{\beta}^{2}{k}^{2}-{\beta}^{2}k-kvc_{{1}}+kxc_{{1}}+k{c_{{
1}}}^{2}-vx}{\beta+c_{{1}}-v+x}}\\ \noalign{\medskip} \frac{1}{ \beta+c_{
{1}}-v+x }&-{\frac {kv-kx-kc_{{1}}+x}{\beta+c_{{1}}-v+x}}
\end {array} \right)
\end{equation}
These matrices are a Lax pair of the following equation
\begin{equation}
(x-v)(u-y)+(c_1-\beta)(x-y)+(c_1+\beta)(u-v)+c_1^2+\beta^2+\beta^2 s=0,
\end{equation}
which is Equation 19 given by Hietarinta and Viallet.
By multiplying the matrix $L_1$ with $2\beta$, taking $c_1=\alpha$ and replacing $\beta^2 s$ with $s$, we obtain the following equation
\begin{equation}
\label{E:19n}
(x-v)(u-y)+(\alpha-\beta)(x-y)+(\alpha+\beta)(u-v)+\alpha^2+\beta^2+ s=0,
\end{equation}
which has the following Lax pair
\begin{equation}
\mathcal{L}_1=
\left( \begin {array}{cc} \beta\,k+u&{\beta}^{2}{k}^{2}-2\,{\beta}^{2
}k-\beta\,ku+\beta\,kx-ks-ux\\ \noalign{\medskip}1&\beta\,k-x
\end {array} \right),
\end{equation}
and $\mathcal{M}_1$ is obtained by replacing $c_1$ with $\alpha$.

Using the gauge matrix
\begin{equation}
G=\left( \begin {array}{cc} 1&-\beta\,k-x\\ \noalign{\medskip}0&1
\end {array} \right),
\end{equation} 
we get a simpler looking Lax pair
\begin{equation} 
\widetilde{\mathcal{L}}_1/k=
\left( \begin {array}{cc} 0&-2\,\beta\,u-2\,{\beta}^{2}-s+2\,\beta\,x
\\ \noalign{\medskip}\frac{1}{k}&2\,\beta\end {array} \right),\ 
\widetilde{\mathcal{M}}_1/k=
\left( \begin {array}{cc} {\frac {\alpha-v+x-\beta}{\beta+\alpha-v+x}
}&\alpha-v+x-\beta\\ \noalign{\medskip}{\frac {1}{k \left( \beta+
\alpha-v+x \right) }}&1\end {array} \right).
\end{equation} 
This Lax pair is useful for calculating integrals for  reduced maps obtained as reductions of equation E19.

The Lax pair $(\widetilde{L},\widetilde{M})$ gives us a conservation law
\begin{equation}
(F,G)=\left(\ln(s+2\beta^2+2\beta(u-x)),\ln(\alpha-\beta-v+x)-\ln(\alpha+\beta-v+x)\right).
\end{equation}
\end{itemize}
\begin{remark}
When $\alpha=\beta$ and $s=-2\beta^2$, equation E19 is a special case of the following equation whose entropy vanishes
\begin{equation}
\label{E:18}
{\rm E18}: \ (p_3v+x)(p_3y+u)+r_3(p_3v+u)=0,
\end{equation}
with $p_3=-1$. This equation is Equation 18 given in \cite{HV}. 
we get a  Lax pair for equation~\eqref{E:18} with $p_3=-1$ and $r_3=2\beta$ as follows
\begin{equation}
\label{E:Lax_18_special}
L=\left( \begin {array}{cc} 0&-2\,\beta\,u+2\,\beta\,x
\\ \noalign{\medskip}\frac{1}{k}&2\,\beta \end {array} \right),\
M=\left( \begin {array}{cc} -{\frac { \left( -x+v \right) }{2\,\beta-v
+x}}&- \left( -x+v \right) \\ \noalign{\medskip} \frac{1}{k\left( 2\,\beta-v+x
 \right)}&1\end {array} \right).
\end{equation}
\end{remark}
%*************************************************************************************************************************************************************************
\section{Conclusion}
In conclusion, this paper has presented for the first time a novel Lax pair for equation E19   given  in \cite{HV}. It reinforces that equation E19 is actually integrable in the sense of having a Lax representation.  This Lax pair was derived by using a double twist of $Q_1$ and taking a limit of this twisted equation.
A double twist  has also been used to derive  multi-parametric versions of the cross ratio equation,   $H_1$  equation and their Lax pairs.  We  would also like find a Lax pair for   equation E18  given by~\eqref{E:18}. We have tried to find its Lax pair from a twist $T(x)=\lambda x$ and a Lax pair~\eqref{E:Lax_18_special}, but the twist matrix does not
exist. Therefore, it leaves an open question whether equation  E18 has a Lax pair. If it does, then how  can one  find  it?

\section*{Acknowledgments}
This research is supported by the Australian Research Council. The author would like to thank Prof Reinout Quispel  and Theodoros Kouloukas for their comments.

\end{document}